%Paper: astro-ph/9508045
%From: mannheim@main.phys.uconn.edu (Philip Mannheim)
%Date: Wed, 9 Aug 95 11:54:38 EDT

\magnification=1000
\baselineskip=0.205truein
\centerline {\bf LINEAR POTENTIALS IN THE CORES OF CLUSTERS OF GALAXIES}
\medskip
\centerline {Philip D. Mannheim}
\medskip
\centerline {Department of Physics, University of Connecticut,
Storrs, CT 06269-3046}
\smallskip
\centerline {e-mail: mannheim@uconnvm.uconn.edu}
\medskip
\centerline{To appear in "Clusters, Lensing,
and the Future of the Universe"}
\medskip
\centerline {Proceedings of the
1995 Astronomical Society of the Pacific Symposium,
 Maryland, June 1995}
\medskip
\centerline{UCONN-95-04~~~~~~~~~~~~~~~~~~~~~~~~~~~~~~~~~~~~August 1995}
\bigskip
\noindent
Abstract:  We make a first application of the linear gravitational
potentials of the conformal gravity theory to the distance scale
associated with clusters of galaxies, with the theory being found
to give a reasonable value for the mean velocity of the virialized
core of the typical Coma cluster, with no need to invoke the
existence of dark matter.
\bigskip
In a recent series of papers (Mannheim 1994 gives a full bibliography) Mannheim
and Kazanas have explored conformal invariant fourth order gravity as a
candidate
alternative to the standard second order Newton-Einstein theory. In their
initial paper they
showed that in this theory the potential
of a gravitational source takes the form $V(r)=-\beta c^2/r+ \gamma c^2r/2$, to
thus both
nicely recover the Newtonian potential and at the same time introduce
a new source-dependent length scale $\gamma$ which would then parameterize any
possible
departures from the standard phenomenology. Moreover, Mannheim (1993) then
showed that
with the use of this potential it is possible to fit the rotation curves of
some typical galaxies without needing to assume the existence of any dark
matter, with the
contribution of the linear potential when integrated over the observed surface
brightness data
then completely replacing the dark matter contribution in the same galaxies.
Intriguingly
it was found from the fitting that the strength of the linear potential of an
entire galaxy
was typically of the
order of the inverse Hubble length to suggest a possible cosmological origin
for $\gamma$. Armed
with an actual magnitude for the galactic $\gamma$ we can now apply the
conformal theory to the
first available distance scale beyond galaxies, namely that of clusters of
galaxies. Interestingly,
we find
that the conformal theory deviates from Newton there by just the amount needed
to nicely accommodate
a virialized cluster core without invoking dark matter, though the theory may
turn out to have some
difficulties should entire clusters prove to be virialized.

In applying the standard cluster virial analysis to theories with rising rather
than falling
potentials two problems arise which are not considered in the standard theory.
First since
the standard discussion generally considers the two-body collision dependent
term in the Boltzmann
equation to be a local perturbation on the global mean field set up by the
gravitational field of
the rest of the galaxies in the cluster we have to reexamine the entire
formalism in light of the
conformal gravity linear potentials which grow
with distance and which can thus not be thought of as producing localizable
collisions at all.
Moreover, in a theory with rising potentials, we are not free to ignore the
effects due to all of
the rest of the galaxies in the Universe, so that even if a system such as a
cluster of galaxies is
geometrically isolated, that does not immediately mean that it is
gravitationally isolated in our
theory or that it is bound purely under its own self forces. As regards the
inapplicability of
the Boltzmann analysis within clusters in the linear potential case, we note
that fortunately
there is an altogether far more general statistical analysis beyond Boltzmann,
namely that based on
the Liouville equation as conveniently treated via the very
general Bogoliubov, Born, Green, Kirkwood, and Yvon (BBGKY) heirarchy (a
heirarchy which reduces to
the Boltzmann equation only under very special circumstances and which remains
valid even when no
such reduction is possible). Since the BBGKY hierarchy is valid for any central
potential it is thus immediately
valid in the linear potential case too,
with our ability to apply the standard Jeans and Vlasov virial equations then
reducing to the
degree to which we can ignore two-particle BBGKY correlations (see Mannheim
1995 for details).
Since the BBGKY relaxation time is currently unknown it is not immediately
clear how much of a
typical cluster has so far virialized (we argue below that possibly only the
core is virialized),
and so we develop below a formalism which can deal with both fully and
partially virialized
clusters. As regards the effect of the global coupling of the cluster to the
rest of the Universe,
this would require the
developing of a theory for the growth of inhomogeneities and galaxy formation,
with the
relevant issue for motions within clusters then being not so much the coupling
of each cluster to
the general Hubble flow produced by a homogeneous background distribution of
linear potential
sources, but rather
its coupling to the deviations from that flow caused by the presence of
inhomogeneities. Since a
theory for the growth of inhomogeneities in conformal gravity has yet to be
developed, we are
currently unable to address this issue explicitly, though as a first step we
will determine
approximately where it is that the cluster begins to merge with the general
cosmological background,
and should thus immediately anticipate that only those regions of the cluster
which lie well within
this merging radius are well enough isolated to be able (if then given enough
time) to decouple and
subsequently virialize.

To treat clusters in general let us consider a spherically symmetric cluster
with matter volume
density $\sigma(r)$ and matter surface density $I(R)$ contained within some
volume of radius
$R_M$, and let us suppose that only some inner kinematic region of the cluster
within some
virialization
radius $r_m$ has so far had time to virialize. Then within the $r <r_m$ region
we may
ignore all two-body correlations with the Jeans equation
$${d \over dr}(\sigma(r)<v_r^2>)+{2 \sigma(r) \over r}(<v_r^2>-<v_{\theta}^2>)
=-\sigma(r)V^{\prime}(r)
\eqno(1)$$
then holding in this region. Despite the fact that Eq. (1) only involves points
with $r < r_m$, we
note that the potential $V(r)$ needed for it is obtained by integrating the
Newtonian and linear potentials over the entire cluster and not just the
virialized region,
with the linear piece (unlike the Newton piece) actually receiving
contributions from points with $r>r_m$. Now observationally
we measure the two-dimensional projected line of sight velocity distribution
average
$<\sigma_p^2(R)>$ which is related to three-dimensional velocity averages (the
ones which appear in
the Jeans equation) according to
$$I(R)<\sigma_p^2(R)>=2\int_R^{R_M} {dr \sigma(r) \over r(r^2-R^2)^{1/2}}
\left(
<v_r^2>(r^2-R^2)+ <v_{\theta}^2>R^2 \right)
\eqno(2)$$
with the integration range all the way to the cluster radius $R_M$ in Eq. (2)
thus involving not
only the desired virialized $r<r_m$ region but also the non-virialized $r>r_m$
region where Eq. (1)
does not hold.
However, integrating Eq. (2) itself over a sphere of radius $r_m$ then yields
(Mannheim 1995)
$$2 \pi \int_{0}^{r_{m}} dR RI(R)<\sigma _p^2(R)>
={4\pi \over 3}  \int_{0}^{r_m}drr^2\sigma(r) \left(<v_r^2>+2<v_{\theta}^2>
\right)
\eqno(3)$$
an exact and completely general relation in which the undesired unvirialized
$r>r_m$ region has crucially been projected out. Since Eq. (3) only involves
the region $r<r_m$,
it thus only involves the region
where we have presupposed the Jeans equation to be valid, with its use then
yielding for the
spatially averaged line of sight velocity
$$2 \pi(\sigma_p^2(r_m))_{av} \int_{0}^{r_m}dR RI(R)=
{4\pi \over 3} \int_0^{r_m}drr^2\sigma(r)rV^{\prime}(r)
\eqno(4)$$
to give finally the form of the virial theorem for partially virialized
systems.

For the most studied cluster, the Coma cluster, the relevant data may be
found in White et al (1993) and references therein. At the distance of Coma
an arc minute is $20/h$ kpc (for a Hubble parameter $H_0 =100h$ km/sec/Mpc), so
that the standard
Abell radius is $75^{\prime}$ for Coma. The surface brightness may be
approximately fitted by a
modified Hubble profile ($\sim 1/(R^2+R_0^2)$) with a core radius
$R_0=9.23^{\prime}$
and number density normalization $\sigma _0 / R_0^3=0.016$ galaxies per cubic
arc min., with the
observed cluster data going out about to $3^{\circ}\simeq 20R_0=R_M$ or so
where we shall thus cut
off the distribution. (A modified Hubble has to be cut off somewhere since it
would
otherwise yield an infinite total mass). White et al (1993) quote a
total blue surface luminosity within the Abell radius of $L_B=1.95 \times
10^{12}/h^2~L_{B\odot}$,
and a mean projected line of sight velocity of 970 km/sec for a convenient
magnitude limited cut on
the data which restricts to $R\leq 120^{\prime}$. For such a mean velocity,
the time required to cross the associated $240^{\prime}$ diameter is $1.5
\times 10^{17}/h$ sec,
which is of order $1/2H_0$, i.e. of order half a Hubble time, and thus we
should not expect the
entire cluster to have yet had time to virialize. Hence, for the purposes of
this study, we shall
simply assume that only the inner region cluster core has so far virialized.
Giving each galaxy an average blue luminosity of
$5.99\times10^{9}/h^2~L_{B\odot}$, then yields
the requisite total $1.95 \times 10^{12}/h^2 ~L_{B\odot}$ surface blue
luminosity within the Abell
radius, to thus fully specify $I(R)$. Using as typical the mass to light ratio
$M/L_B=5.6hM_{\odot}/L_{B\odot}$ obtained for the galaxy NGC 3198 in
Mannheim(1993)
enables us to determine the mass
volume density associated with $\sigma (r)$. It is very convenient to express
this mass density in
units of the standard critical density $\rho_c=3H_0^2/8\pi G$, and we find that
$\sigma (0^{\prime})=241.5\rho_c$, $\sigma (56.8^{\prime})=\rho_c$,
$\sigma (120^{\prime})=0.11\rho_c$, and $\sigma (185^{\prime})=0.03\rho_c$. The
cluster is thus
apparently merging with the general cosmological background at no more than
$185^{\prime}$ or so,
and would be restricted to the first $57^{\prime}$ if the density of the
Universe is critical. Thus
in a low density Universe we would put the edge of the cluster at
$185^{\prime}$, while in a high
density one we would only consider the potentials of the first $57^{\prime}$ of
data as contributing
to the velocity dispersion, with the next $128^{\prime}$ of data then only
contributing along with
the rest of the galaxies in the Universe to the general Hubble flow. (Noting
that the conventional
estimation of the cosmological ratio $\rho/\rho_c$ is made in comoving
coordinates while our
analysis here involves the same ratio in static coordinates, our determination
of where the static
cluster actually merges with the comoving background is thus perforce only a
rough estimate.)
Since the actual density of the Universe represents one of the key unknown
issues in cosmology, we
shall calculate cluster core virial velocities for both the high and low
density Universe cases, and
actually find below that the values that we then obtain turn out to be
insensitive to where the
cluster ends. Moreover, no matter where we put the edge of the cluster, we
should not expect the
matter within the cluster but close to this edge to necessarily be completely
decoupled from the
cosmological background, so that again only core virialization seems
reasonable. However, since the
modified Hubble is still a falling profile, it turns out that one quarter by
volume of the entire
$20R_0$ cluster is contained within $r\leq 2.5R_0$, while one half is contained
within $r\leq 5R_0$. Thus virialization of only the core region is quite
non-trivial.

Taking $V(r)=-\beta_{gal} c^2/r+ \gamma_{gal} c^2r/2$ as the potential put out
by a typical
individual galaxy and taking $N$ to be the total number of galaxies contained
in the entire cluster
(and not just the number
$N(r_m)$ contained in the virialized $r\leq r_m$ region) fixes the overall
normalization
$(N\beta_{gal} c^2/R_0)^{1/2}$ of the Newtonian potential contribution
to the cluster virial, while taking as typical the NGC 3198 gamma to light
ratio of
$9.2 \times 10^{-40}h^3 /cm/L_{B\odot}$ also obtained in Mannheim(1993) then
enables us to fix
the overall normalization $(N\gamma_{gal} c^2R_0)^{1/2}$
of the linear potential contribution as well. Thus for a Coma cluster composed
solely
of luminous matter alone, the overall normalizations $(N\gamma_{gal}
c^2R_0)^{1/2}$ and
$(N\beta_{gal} c^2/R_0)^{1/2}$ needed for Eq. (4) take respective values of
10960 km/sec and 576
km/sec for a cluster cut off at $R_M=20R_0$ ($N=425$ galaxies).
In the absence of any dark matter the luminous Newtonian
contribution to the virial is thus negligible, while the linear contribution
associated with the luminous matter is substantial. Specifically, if the entire
$R_M=20R_0$ cluster
is virialized,  Eq. (4) yields a virial velocity $\sigma_p(20R_0)=10178$
km/sec, while also yielding
partial virial velocities $\sigma_p(R_0)=1089$ km/sec, $\sigma_p(1.5R_0)=1678$
km/sec,
$\sigma_p(2R_0)=2195$ km/sec, and $\sigma_p(6.15R_0)=5018$ km/sec in various
inner regions.
Similarly, if we cut off the cluster at $57^{\prime}=6.15R_0$ (to yield $N=242$
galaxies,
$(N\gamma_{gal} c^2R_0)^{1/2}=8261$ km/sec, $(N\beta_{gal} c^2/R_0)^{1/2}=435$
km/sec)
we obtain the partial virial
velocities $\sigma_p(R_0)=1028$ km/sec, $\sigma_p(1.5R_0)=1583$ km/sec,
$\sigma_p(2R_0)=2070$ km/sec,
and $\sigma_p(6.15R_0)=4885$ km/sec. The core region velocities are thus
essentially insensitive to
whether we use a high or low density Universe cut-off. From the line of sight
velocity data points
we find that
the numerical average of all the points in $R\leq 1.3R_0$ is $1200\pm 195$
km/sec, while
that of the all the points in $R\leq 1.7R_0$ is $1185\pm 195$ km/sec. However,
before we assess the
significance of these numbers, it is important to note that once less than the
entire spherical
cluster is virialized, then any given line of sight through the sphere, even
those at small impact
parameter $R$, will pass through both virialized and non-virialized regions
(the integration in
Eq. (2) is to $R_M$ and not merely to $r_{m}$), so that the detected projected
velocity at that $R$ will include
some non-virialized contributions as well. For instance, if $r \leq 2.5R_0$ is
virialized, then out
of a $20R_0$ cluster the percentage of line of sight material which involves
unvirialized radii
$r>2.5R_0$ is $25\%$ at $R=1.5R_0$, $44\%$ at $R=2R_0$, and of course $100\%$
at $R=2.5R_0$. Thus
even though the partial virial of Eq. (4) itself only involves integrating up
to $r_{m}$,
the very use of initial input raw velocity data to estimate a magnitude for a
virialized
$\sigma_p(r_m)$ in the
$r<r_{m}$ region becomes suspect once the cluster is less than fully
virialized. From our calculated virial
velocities we see that
conformal gravity would thus appear to have no difficulty accommodating a
virialized inner cluster
region of the order of $r_{m}\sim 1.5R_0$ without needing to invoke dark
matter, and given the just
noted limitation on the use of the input velocity data in partially virialized
systems, the theory
could possibly even accommodate up to $r_{m}\sim 2.5R_0$, a region which
contains close to one
quarter by volume of all of the matter in the entire $185^{\prime}$ of the
cluster. Moreover, given
the relevant time scales which were discussed above, it would even appear to be
quite reasonable to
expect inner region virialization up to one or two scale lengths or so. While
we would certainly
not expect any larger a portion of the cluster
to have yet virialized, a first principles determination of the two-body
correlation function and of
its potential impact on Eqs. (1) and (4) could nonetheless prove to be very
instructive,
and might possibly even turn out to be definitive for the theory. (It is also
possible to test the
conformal theory in a way which is actually insensitive to how big a fraction
of the cluster has
in fact virialized, viz. cluster gravitational lensing which responds to all
the matter in the
cluster virialized or not; thus a yet to be made study of the conformal theory
predictions for
lensing should eventually provide an independent and definitive way of testing
the theory on whole
cluster scales.) Other than this issue though, it would appear that, in the
first instance at least,
the conformal gravity theory is indeed capable of meeting the demands of
cluster virial velocity
data, with the linear potential theory thus readily being extendable from
galactic scales up
to the much larger ones associated with the virialized core regions of clusters
of galaxies without
encountering any major difficulty. This work has been supported in part
by the Department of Energy under grant No. DE-FG02-92ER40716.00.

\smallskip
\noindent Mannheim, P. D. 1993, ApJ, 419, 150.
\smallskip
\noindent Mannheim, P. D. 1994, Foundations of Physics, 24, 487.
\smallskip
\noindent Mannheim, P. D. 1995, {\it Linear Potentials in Galaxies and Clusters
of Galaxies},
preprint UCONN-95-02, March 1995.
\smallskip
\noindent White, S. D. M., Navarro, J. F., Evrard, A. E., and Frenk, C. S.
1993,
Nature, 366, 429.

\end